**Resolution of Kantor and Babcock-Bergman Emission Theory Anomalies.**

The SR postulates emerge from SMOKE, mirrors and a quantum mechanism.

Peter A Jackson[1]  and  John S. Minkowski.[2]  V3c

*[1]RAS. AAS. RIBA. Orchard House, Whitstable, Canterbury Kent CT5 3EX, UK,
peter.jackson53@ymail.com  [2]Baltimore, Maryland, USA. jminkow1@jhmi.edu*

## Abstract

Kantor's 1962 interferometer result supporting the emission theory of light was tested by Babcock and Bergman (B&B) in 1964, but with rotating glass plates placed in a vacuum. The result was different and consistent with Einstein's Special Theory of Relativity (STR). Two anomalies remained "*not fully understood*"; A 0.02 fringe shift was still found (Kantor found 2.9), and the kinetic fringe shift with rotating plates was far *smaller* than the shift with the plates static. The results falsified the theory that light passing out of the glass continued at c+v in the lab frame, but the anomalies were not resolved. We review these alongside related and poorly understood effects including kinetic reverse refraction and non-linear optics. We also consider advances in science and astrophysics and find and describe a theoretical resolution. We find that Kantor's finding may also be '*apparent*' without violating the postulates of the STR or invoking an absolute 'ether' frame. Relationships between Maxwell's near field transition zone, photo-ionization, non-linear optics and the surface electro/magneto-optic Kerr effects emerge, building an ontological construction which we describe and quantify. Proper Time is found to be required for Proper speed. A relativistic theoretical model is built from a diffractive mechanism, as used by the US Naval Observatory for accuracy in the AA2010 aberration declination model, but now with a "*consistent relativistic theory*" to support it. We show how the inconsistency between Michelson's 1924 finding supporting ether and his famous 'null' result may arise from systemic '...asymmetry of crossing counts. The quantum mechanisms of Raman (and coherent forward) scattering, optical axis rotation and CMB frames last scattered are shown to directly derive the STR postulates. Some further experiments are suggested.

## 1. Introduction

Wallace Kantor (1962) reported experimental findings that "*the relative speed of light is dependent on the uniform motion of the source.*" which contradicted the postulates of the STR. Plates of glass on a rotating arm were placed between the static mirrors of a modified Michelson-Morley (M&M) interferometer moving in free air in the lab frame. Kantor found that light passing through the glass was re-emitted at c with respect to the glass and continued through the medium at c in the frame of the re-emitting glass surface. In 1964 Babcock and Bergman (B&B) repeated the experiment with four times greater sensitivity, using a reversible motor, and measuring; "*the speed of light which has passed through moving glass plates by the observation of the shift of interference fringes*" in a vacuum, and "*with the fringes observed at infinity.*" Where Kantor's emission theory predicted 2.9 fringes B&B found "*less than 0.02 fringe.*" The B&B result was consistent with the propagation speed of c/n in the inertial frame of the glass plates, and c = d/t while passing through the vacuum, consistent with the postulates of the STR. The result was also in accord with the proof of the first STR postulate using binary stars, where the light travels at c irrespective of the approach of recession of the star as first found by Comstock (1910) and de Sitter (1913).

J.G. Fox suggested (1965) Ewald-Oseen extinction at change of propagation velocity after emission from the ions and electrons of a mirror at c + $v$ in the laboratory frame. The velocity of radiation "*received by the air molecules as a frequency of* $\nu(1+\beta)(1+\beta) \approx \nu(1+2\beta)$... *is quickly localized by them to* c [$\lambda/(n-1)$ = 0.3 mm (in air) at NTP]," (Fox 1965), where $v \equiv \beta c$ is the velocity of the glass and $\nu$ the frequency of the light source. Kantor considered experiments inconclusive, extinction unproven and said; "*Rather than confirming the absolute speed of light, these null result experiments can also be regarded as merely showing the obliteration of the relative speed of light as it propagates through various media.*" (Kantor 1971). Of course Kantor had a logical point, but any implication that c + v would then occur in a vacuum was still contrary to de Sitter's proof of the first STR postulate, consistently confirmed by spectroscopy including by Brecher (1977) who reasoned that extinction could not apply over short vacuum distances. So the problem then remained; ***How could extinction occur in a vacuum? Or what else was happening?***

Speed, as distance over time d/t, requires a datum; two frames, an emitter frame and background reference frame, or two reference points (a known distance) to be assignable. The background 'ether' frame had to be removed to allow the STR to explain constancy of c (CSL) for all moving *receivers*. But this then allowed only relative velocity between bodies. Einstein's tried (Leiden 1921<) to reinstate an ether frame to defeat emission theory, (c always with respect to emitter) which confounded c = d/t. But such a background or 'absolute' frame then confounded any explanation of CSL for *observers* moving through the ether. Length contraction had been invoked to escape that apparent paradox, but has so far still never been proven so theories of speed c in vacua with respect to the original emitter have endured. The B&B experiment might have been conclusive but the "*not fully understood*" fringe shifts and full theoretical basis remained unexplained. To explain these we invoke related effects and phenomena, some not fully understood and some not currently considered related. These include among others Maxwell's near and far field transition zone, non-linear optics, the fine structure constant, magnetohydrodynamics, Raman scattering, aberration, birefringence, surface charge, plasmons and the Surface Magneto Optic Kerr effect (SMOKE). Light as waves overcomes the problem of photon momentum in media, avoiding the Minkowski/Abraham tensor issue and even suggesting the Einstein-Loeb model may be locally valid.

## 2. Birefringence and Extinction

First found in calcite in 1669 birefringence is considered as decomposition of a 'ray' of light into two rays. It may be more properly considered in terms of plane waves and the 'optical axis'. Raman found birefringence (1921) in diffuse gas in his 1930 Noble prize winning work on scattering, and free electron/ proton plasma of highly birefringent with a refractive index n =~1. We know that the quantum vacuum is also birefringent. We suggest that kinetic effects related to twin 'light paths' (refracted axis) are evidence of gradual extinction of emissions due to coupling interaction in a diffuse particle medium over vast distances. Laplace considered 'intensity' in r = (mass) *M sin z.* But consistent with Raman's findings in diffuse media we predict that the two optical axes also represent different *velocities* of propagation, as c in both the 'arrival' and re-scattered frames. We propose a series of experiments to confirm this theory including *[Exp.1]* with a plasma particle chamber and varying emission rates in an interferometer to compare arrival times of the two (birefringent) components of light, which we propose arises from a relative rotation of optical signal axis.

Enders (2011) suggests that wave extinction at the surface of a new medium contradicts the Huygens-Fresnel Principle (HFP) which says that the incident wave is absent after excitement of the secondary wave source. We suggest that extinction will be progressive and subject to particle density so both waves will exist during the process. Neither Birefringence nor extinction conflict with the HFP, but neither do they provide additional clarification. Some confusion exists between the axis of polarity and the optical axis. The former relates to Faraday rotation of polarity *around* the optical axis or axial vector, the optical axis giving the apparent source position, normally considered as the 'light path vector' in geometrical optics. In a solid such as calcite the optical axis is considered as a function of the medium particles. In plasma the change in optical axis is considered here as having two components. The first is the normal charge/scattering dispersion delay giving refraction, which is itself still not fully understood. We propose that a second change is kinetic, due to the axis of the arriving signal waves being rotated on re-emission by charge asymmetry due to lateral motion of the particles, or bulk plasma flow during the non zero charge time. This rotation of re-emission due to *asymmetry of charge* is additional to the more fundamental effect arising from the first STR postulate, where the signal is 'carried' by the new medium if in relative motion.

Recent findings from the *SAURON* survey (Davis, 2001) and the ensuing ATLAS[3D] results (Emsellem, et al. 2011) include spectroscopic red and blue shift derived halo rotation velocities. These findings are consistent with the well known kinetic Sunyaev-Zeldovich (kSZ) effect but are troublesome without invoking refraction of light in 'space', *apparently* contradicting the STR. We find a consistent explanation by including scattering, birefringence and gradual extinction in diffuse media, as found with radio signals (Sukamar 1987). The vacuum itself is considered as a medium, but an exceptionally diffuse medium, implementing change very gradually across exceptional distances measured in kilo-parsecs. Vacuum extinction cannot then be detected in a laboratory, except by using partial vacua. We propose a partially evacuated chamber at varying density as the basis for further experimentation *[Exp.2]* but more thought is still required. An important experiment is R. Wang's linear 'Sagnac' interferometer (Wang 2006). When  using a fully evacuated wave-guide tube, whatever the frame of the emitter, a glass sealing plate or the walls of a tightly wound tube will re-emit the light at c in the frame of the tube. This may explain why propagation in this case remains at c in the wave-guide frame.

### 3. B&B Accommodation.

There does now seem to be a quantum based solution possible for the outstanding anomalies of the Babcock and Bergman findings. The speed, total thickness and refractive index of the glass and Fresnel drag co-efficient were applied, and no systemic errors were apparent (see Babcock-Bergman 1964). We must first remind ourselves the Fresnel's  refractive index 'n' may only be found by experimentation, but is well known for glass, although only approximated by B&B at n~1.5. When determining refractive index, however, it is common to just consider a single piece of glass of a given thickness. This procedure does not allow for surface absorption and emission delays, which are not yet well understood, but which we predict will be found approximately equivalent to the surface losses to transmission of heat. In the case of heat the transmission loss (k value) is an additional ~0.18k 'per surface' in addition to the transmission loss from the glass itself,  making a significant contribution to total loss. Two separate glass plates each 0.35cm will then have greater transmission loss than one sheet 0.7cm thick. Experiments with light passing through small apertures have shown transmission delays due to surface oscillator interactions (Dogariu et al. 2001). We suggest an experiment *[Exp.3]* comparing varying numbers of surfaces for a given glass thickness to quantify the effect. We predict that the result will be found to give in the order of the additional 0.01fringe shift which B&B found as the unexplained fringe shift with the two glass plates stationary so with no Fresnel flow.

The question of the absorption and emission processes at the surface is again raised, and we must consider the additional time delay as well as the changes in transmission velocity between the media. Firstly however we must consider the interesting additional anomaly of the low magnitude fringe shift derived solely from the velocity of the glass plates. The relative speed of rotation (allowing for changing glass vector due to rotation) was an average of ~90 rps. The additional fringe shift change in this case over the static case was ave ~ 0.004 fringes. We suggest a kinetic solution based on the phenomena considered in section 2 above. B&B gave; v(glass) /c = 1.27 x $10^{-7}$ a very small percentage of the speed of light. We clarify that the refractive index of a medium is a constant irrespective of any relative motion of the medium (representing an inertial frame) and its surroundings. The effect of the n value of glass of ~n = 1.55 will then give a reduction to speed of propagation in the mirror frame of ~1/1.55 = ~200,000 km/sec. This change applies to both the static and rotating case, and will be significantly greater than just the kinetic effect of the moving glass. The latter gives only the 0.004 fringe shift, or ~1/5th of the total static shift including any 'surface transmission' losses.

It is apparent that the 0.004 kinetic shift is slightly higher than might be expected rather than anomalously low. Even a good vacuum on Earth contains a significant particle density, as does the Intra Galactic Medium, but at entirely insignificant densities. B&B reported fringe instability due to air turbulence prior to evacuation, and ran the tests at a medium vacuum of ~0.02 Torr where a significant particle density of ~1.5kPa remains. Particle kinetics will then be likely to have an effect on the result and a lower figure should be achievable in a higher vacuum. Whether or not this is the case we are now able to explain the B&B results. The Kantor result, if found in a significant depth of air circulated by the rotating glass plates, would certainly also be possible if the kinetic shift effect were thus increased by a factor of 7 from ~ 0.04 to ~ 0.29. We have, in any case, now identified three separate but related causes of the speed change of light passing through the glass when observed from the laboratory frame, all apparently consistent with the STR and also giving results consistent with the STR. None of these are fully understood or applied within current science;

1. *Kinetic.* Due to the change in position of the glass during the time taken by the light to pass through it.
2. *Medium.* By c/n due to the thickness and refractive index n of the glass, only found experimentally.
3. *Surface*. Absorption/transmission losses at speed change interfaces, normally included in 2 above. *(Addendum; ~28-57 fs. per surface subject to relative polarity).*

The first two are the more familiar, though often not properly considered independently. We will theoretically consider the third as it is not normally identified as the independent effect witnessed by the B&B results. First we discuss reverse refraction, a kinetic optical effect also poorly understood, rarely considered and not yet assimilated into theory. We now allow assimilation.

## 4. Kinetic Reverse Refraction

This well known geometrical optical effect is found experimentally between co-moving media and apparently violates Snell's Law of Refraction (Ko, Chuang 1977, Mackay, Lakhtakia 2006). When observed from an incident frame, light at near normal incidence passing into a co-moving medium appears to be 'dragged' by the medium (Grzegorczyk 2006). However, when observed from the frame of the second medium it is found that the optical axis has actually reversed, and the 'light path' is refracted back towards the normal. When considered in terms of quantum optics this reversal highlights the important fact that the optical axis of light is *rotated* at the interface of a co-moving medium. We are also reminded that 'vectors' within another inertial frame are not 'real' but are an *effect* which is subject to the observer's frame and the rules of 'Proper Time'. Speed and direction are only valid using Proper Time, which is time in the frame of the observer, giving *propagation speed*. '*Co-ordinate time*' only gives '*apparent*' speed. Detecting by direct interaction is a different case to 'observing' via secondary emissions, with different results. A 'Ray' and its 'light path' are then inadequate concepts. For consistency we must consider plane wave propagation and the local 'optical axis' of scattered light. The rotation then informs us as to the mechanism at the medium interface.

Stellar aberration has also remained problematic because the phenomena of kinetic reverse refraction was not known by Lodge in 1893 when interpreting the effects of light entering a spinning glass disc. Observed correctly from the *disc* frame (Lodge wrongly used his Lab to represent the Geocentric frame) the optical axis is rotated *towards* the normal, as found, but now explained using only EM waves and without invoking ballistic photons or ether. A full relativistic theory for aberration then emerges consistent with observation and the empirical diffraction addition currently required for precise predictions. The troublesome Aberration 'Constant' was abandoned by the IAU in 2000 due to inconsistencies. The US Naval Observatory developed the SAM13 Ionospheric model but have to add refraction empirically to NOVAS algorithms for accurate Astronomical Almanac (AA2010) declination aberration data. USNO circular 179 (Kaplan 2005) admits that "*a consistent relativistic theory of Earth rotation is still some years away.*" At 60º-90º zenith the significant corrections 'by hand' to algorithmic predictions are <34 arc minutes ($\lambda$ ~0.5 microns). We now offer a consistent theoretical solution. Magnetohydrodynamic turbulence and incomplete extinction ensure that a <5% inaccuracy (Heymens et al. 2005) of aberration remains (see Part 8). The atmosphere's rotation within the ***non***-rotating ECI frame is part of the solution, giving turbulence, ducts and 'green flashes' from temperature inversions. An intuitive cause of red shift of evening sunlight emerges 'scattered' from protons and dust receding from the sun and ECI shock frame with planetary rotation. A fluid dynamic coupling effect at the shock, where the ions on both sides re-emit at the local c with $\Delta\lambda$, resolves the balance. *(Addm. See the standard two-fluid plasma model; Shumak et al. 2004)*

In finding consistency with STR's postulates we then imply that the light speed reverts to c with respect to the 'vacuum' on passing from glass to vacuum both by the factor c/n *and* the kinetic factor v of the relative media motion. The latter may logically imply a vacuum inertial frame, or 'ether', so the apparent paradox of the STR is so far unresolved at the quantum level. A possible solution emerges from the medium interface mechanism, consistent with the reference frame by which the speed of a proton bunch through an accelerator vacuum tube is measured. This is considered as the frame of the static electromagnetic (EM) field of the magnets used to accelerate the bunch in the tube. It is also the frame of the static part of the 'virtual electron' cloud which builds in the tube with increasing bunch velocity, consistent with the Unruh effect (Crispino et al 2008). In invoking this effect for 'glass plate' shaped particle bunches moving through vacua we next consider less familiar aspects of Maxwell's equations.

## 5. Maxwell's Near and Far Field Transition Zone.
The behaviour of EM waves at the near and far field transition zone (TZ) of an emitter is very poorly understood. Six radio engineers familiar with antenna science asked for the equation for the TZ position, may give six different

'precise' answers. For wavelength λ answers may range from $\lambda/2\pi$ to $5\lambda/2\pi$ or 50D to $2/\lambda$ to $2D^2/\lambda$ where D is the transmitter dish radius or antenna length. For visible light this (Fraunhofer) transition distance from the surface of a small mass at rest is ~1 micron<1mm. Surface phenomena normally considered unrelated include surface charge, static electricity, fine structure electrons, atomic coupling, plasmons, the outer limit of Fresnel (and inner limit of Fraunhofer) diffraction, and the domain of the laws of geometrical optics. The near field is referred to as the quasi-stationary, 'static', reactive and extinction zone, where "*scattering intensity hot spots*" (and) "*striking differences in the phase functions*" (Quinten 2007) can occur. The TZ, also termed the induction field, is turbulent and dynamic with high plasma energies. HFP has been found valid within it (Depasse et al. 1995) implying extinction. The far field coupling regime at $>10^{-6}$m is also termed the radiative field or radiation zone (Lamothe et al. 2007). At radio frequencies the far field is where stable reception starts.

We need to find a conceptual ontology to derive the observed classical effects before applying the normal mathematical tests, so we first consider logic and phenomenological aspects. We find we must reconsider our limited knowledge of the quantum processes at the zone where light propagation speed increases from ~200,000 to 300,000kps on entering air from glass. Light achieves this *speed change* even before we consider the *additional* 'kinetic' speed change due to relative motion of the media (see part 1). The Dynamic Casimir effect shows electron production at moving surfaces, but then gives c in the *medium* frame not the emitter's supporting the hypothesis of fluid dynamic coupling at the TZ. 'Coherent forward scattering' in plasma is invoked as the key TZ mechanism.

Our knowledge is limited and we have incomplete understanding of kinetic and refractive index derived speed change processes at nanoscale level. There are obvious practical difficulties in carrying out experiments at fast moving media interfaces, but a number of interesting and incidental effects have been found. We identify some relevant effects and consider their implications. In particular we review knowledge of the population of the near field and transition zones, some strange behaviours and effects on EM waves. We thus briefly consider the Kerr effects, including the surface magneto optic Kerr effect (SMOKE), surface plasmons and the strange world of non-linear optics.

## 6. Surface Plasmon Resonance (SPR) and SMOKE.

A plasmon is considered as a nanoscale quanta of plasma oscillation, as free electrons on the surface of a solid charged by light (Hecht et al. 1996). SPR is collective oscillation, resonant with the wavelength λ. Plasmons are considered as 'quasiparticles', at media interfaces (usually studied at metal and dielectric media) interacting strongly with light, holding high surface energies and electric fields. Plasmons are closely associated with the bizarre effects of 'non linear optics' (NLO) at high light intensities, quantum cavity opto-mechanics and plasma Surface Enhanced Raman Scattering (SERS) (Kneipp 1997). NLO includes many poorly understood phenomena such as the second, third and higher harmonic generation, creating frequencies from twice the original to ~<1000 times greater. X-ray frequencies may be produced from visible light in this way. Plasma 2-fluid motion (*Shumak 2004*) then implements Δλ and local c

Superposition of waves also fails in the domain of NLO, and 'self phase modulation' (SPM) is caused by temporal intensity variations which give temporal variation of the refractive index n. *Spatial* variation of the refractive index also arises from the electric and complex Surface Magneto-Optic Kerr Effect (SMOKE) where the optical and polarisation axes are rotated, which also produces birefringence. As has been pointed out by Qui and Bader (2000) the Kerr effects are far from understood macroscopically or theoretically complete. In the 'DC Kerr' effect the index value n is changed by the square of a changing electric field. The Kerr constant for a medium $K$ may vary between ~$10^{-22}$ and $10^{-14}$ m$^2$/v$^2$. Where the electric field strength is $E$; $\Delta n = \lambda K E^2$. SERS is attributed to Surface Plasmon Resonance but transmission enhancements of >7 as well as suppression have been more specifically shown to be modulated by wave diffraction from 'sub-wavelength surface features' (Lazec, Thio 2004), which would be consistent with a transition zone velocity change. We propose that all these effects may be rationally explained by a better understanding of propagation kinetics. In the AC or Optical Kerr effect the light itself produces the electric field and Δn giving unstable self-focussing/phase modulation. The result is an *intensity dependent refractive index* (IDRI) limited only by high 'multiphoton ionization' or the more familiar *photo-ionisation,* effectively ion production from light energy *(Addm.; Higuchi et al Jan. 2013 PRA 87 013808).*

Two-fluid surface plasma spatial distribution is dynamic. At energy density limit ~$10^{20}$-$10^{22}$ electrons/cm$^3$ optical breakdown (OB) mode causes shock waves and thermal damage. *The Lorentz Transformation is then implemented by **wavelength limit gamma** approaching OB.* At lowest energies light may be 'frozen' and re-released as demonstrated by Harvard's Hau Lab in Bose-Einstein condensate. With increasing light intensity and 'quantization' into plasmons, the ionization rate and the electric field also increase. It is in this spatial zone that the wavelength dependant and poorly understood dynamic 'non-linear' optical effects found. It is clear that many of these non-linear effects, or violations, are closely equivalent to those of Maxwell's transition zone (TZ) at the Fraunhofer distance including the breakdown of geometric optics, Fresnel Refraction and also of Snell's Law where media co-motion exists. The solution applicable to kinetic reverse refraction is valid; which is an observer acceleration to a new frame, so *recovering c and the laws of geometrical optics* in the new frame. TZ positions for emitter diameter D are commonly considered as 2D$^2$/λ. We must, however, remember that the distance from 'surface' to TZ and from emitting TZ *particles* to the Fresnel limit may prove to differ. As all signal wavelengths dictating Fresnel limits 'exist' at once, then TZ positions will also vary widely.

The STR and B&B result are consistent with the apparently anomalous effect of light reflected from a mirror moving in a vacuum reflecting at c reciprocal to the incident velocity, not c with respect to the mirror. This counter intuitive result was as found by Michelson in his 1913 experiments with rotating mirrors. The effect may now, however, be logically explained via the proposed scattering mechanism. Whether considering a polished metal surface or silvered glass, the TZ would implement the absorption and kinetic change by re-emission. The incident (outer layer) photo-ionised particles are the new reference for the emitted speed c. We suggest that this effect alone is adequately compelling to justify significant experimental falsification.

## 7. Lateral Motion.
How does relative lateral motion of the media give Kinetic Reverse Refraction? If coupling with particles at rest gives symmetrical charge then we suggest that lateral relative motion results in asymmetry of charge. The change of apparent

'light path' termed KRR can then be more consistently characterised as rotation of the optical axis (apparent source position) when re-emitted, accompanying Faraday rotation of polarity. In the case of an asymmetric charge, a mechanism producing ellipticity and elliptical polarity also emerges. The effect is a small angular deflection of the re-emission axis of light from the 'moving' scattering particle. This optical axis rotation due to charging asymmetry now allows KRR to be assimilated into theory, including as the source of wave based stellar aberration. Faraday rotation of polarity and ellipticity are the natural results (see Fig.1). The role of Huygens construction is confirmed by the self focussing and 'regenerative' quality of a Bessel beam when its 'path' has been blocked.

Unlike the ballistic model, the mechanism invoking optical axis rotation consistently applies to waves, and also preserves the *integrity of the causal wavefront*. The translation effect across the whole refractive plane derives the departure of the optical axis from the wavefront normal. Circular polarity then becomes elliptical. Bi/multirefringence in plasma then naturally emerges from this 'progressive' interaction process. The non-zero interaction time giving rise to the axial rotation was recognised by Max Planck in his 'New radiation hypothesis' (1911). The energy structure of his thesis was not adopted, so the implications of the non-zero charge time were not studied or the consequences analysed. Figure 1 shows some of the effects of a circularly polarised emission or re-emission (or 'a photon') translating downwards via interactions with the particles of the new refractive plane dielectric medium *with lateral media co-motion*. Note the ellipticity, axis change, and conservation of the causal plane.

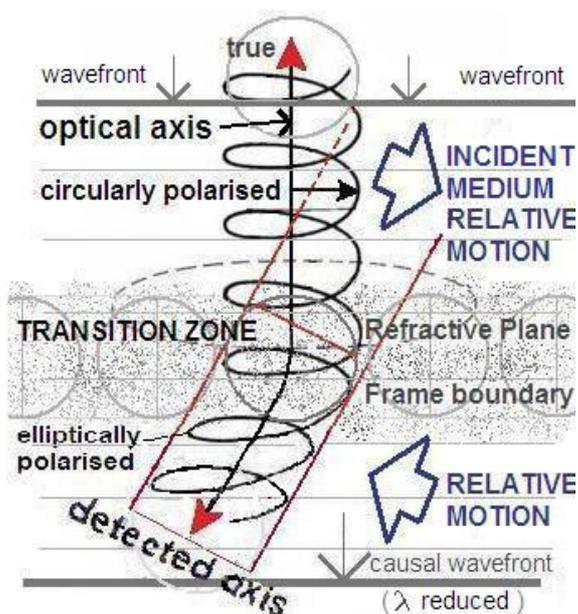

*Figure 1. Transition Zone scattering, giving non-linear optics effects at electron shocks. The 'Optical axis' rotates, wavelength changes (due to relative v) and polarity is ellipticised. The causal wavefront is conserved as the optical axis is NOT perpendicular.*

We suggest that Fresnel refraction may be recovered beyond the Fraunhofer distance and the TZ and in the same way that Snell's Law is recoverable in kinetic reverse refraction. This is by an *observer* acceleration into the new inertial frame, so only when using *Proper Time*. We're also therefore suggesting that non-linear optical effects may then be recovered, which *recovers c and the laws of physics* in the new inertial frame. The quantum mechanism invoked, based on Raman atomic scattering will then be consistent with the postulates of the STR. We also point out that the mechanism is reciprocal so is also fully symmetrical. If this proposed ontological construction of phenomenological relationships is experimentally verified then a *direct causal relation* between quantum mechanisms and Relativity will have been be identified. Altewischer et al. (2002) have already discussed the compatible quantum and macroscopic nature of surface plasmons; "*in the sense that they involve some $10^{10}$ electrons*" (and) "*because they can act as intermediates in transmitting entangled photons to yield the expected fourth-order interference*". We point out that the mechanisms identified lead to a more consistent kinetic derivation of relativity and classical physics direct from the quantum mechanisms. Figure 2 shows the observed effects from the evolution of interaction through the TZ of a mirror in motion in a medium.

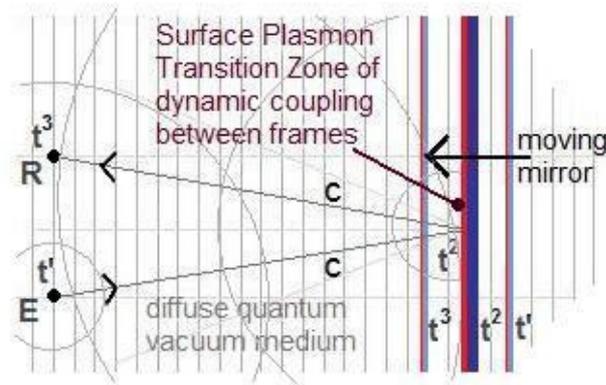

*Figure 1: Mirror in motion during crossing of Transition Zone (TZ). Light propagates at c with respect to the 'far field' medium outside the TZ's two-part 'dynamic coupling' region. PJ 2012.*

## 8. Incidental effects

The limited understanding of non-linear optical effects may be improved by the consideration of a change in wave propagation speed consummate with $\Delta\lambda$. Such 'dispersive' effects are of course found in refraction, and we postulate are modulated at the transition zone (TZ). For instance it is found (Bloembergen 1965 [1977]) that the speed of light in a non-linear medium depends on light *intensity*. We briefly identify other effects which may bear some relation when looked upon from this new perspective. As the intensity of incident light also dictates TZ particle density, then a laser beam might simulate the effects of co-motion between media in a vacuum. Lasers are indeed the normal and an effective means of promoting plasmons. We thus propose an experiment similar to that of B&B in a vacuum *[Exp.4]* but using laser light of varying intensity to find and assess the kinetic effects on activity. The less simple task of moving the emitter and/or the surface plane in the vacuum should allow more to be learned. We predict that electron densities increase along with frequency in proportion to the incident plane velocity in a vacuum, increasing the kinetic optical effects. This may support the notion that the most diffuse vacuum medium, or at least any vacuum containing an em field, may nonetheless be considered a medium, with optical effects ~proportional to particle density.

In considering the optical breakdown limit of electron density we find an analogy with meteors and spacecraft re-entries where the peak plasma density is set well off the surface itself, and also blacks out em transmission. The shock frequencies exceed $10^9$cps at >100km altitude (NASA) in the planetary ionosphere, driven by ion collision intensity. The International Standard Plasmasphere-Ionosphere model (SPIM) uses effective altitudes of <20,000km and <1,000km and maximum vertical total electron content (TEC) electron density may be ~ <40 x $10^{16}$m$^{-2}$, with flux in the lower bands peaking at over three times the density at 850 km. (NWRA). The heliosheath is estimated, from *Voyager* and *SOHO* data (Czechowski 2006) at a flux of ~1.13 x $10^{14}$cm$^{-2}$ at the termination shock of the solar winds against the ISM. Pair production from the vacuum field and increased particle density and frequency with velocity are consistent with the propagation of virtual electrons in accelerators, also the kinetic astrophysical effects identified in part 2, and findings of peak propagation at relative velocity peaks in galaxies (Wozniac, 2007, Dobbs, Pringle 2010). The localisation of velocity suggested by Fox from extinction via coupling and scattering emission at local c does have a philosophical kinetic equivalence to the 'local reality' of inertial frames which Einstein described long before space exploration. We explore the ontology of this equivalence and find Ewald-Oseen Extinction consistent as a concentrated boundary condition effect from Maxwell's equations, which is as found by Dialetis (1978).

## 9. Magnetohydrodynamics

We invoke particle magnetohydrodynamics at the transmission zone (TZ) which gives a kinetic relationship with the Navier-Stokes fluid dynamic mixing process familiar at the ionospheric bow shocks of planets and stars. The vacuum particle densities at bow shocks are measured at $<10^{14}$cm$^3$. The particles on the near field side of the zone or shock are in the inertial frames of the moving body, and those photo-ionised at the incident (far field) side are in the incident frame. The severe turbulence arises from the constant hydrodynamic particle mixing process. In atomic scattering Raman (1921) found that electrons in a 'turbid medium' re-emitted light at c with respect to the state of motion (frame) of the local bulk plasma. This seems unsurprising, however, when we consider the greater implications we find that we are then discussing two separate local speeds c, or rather c with respect to two particle scattering surfaces in different states of motion. This change in relative speed is indeed what is found between the Earth centred inertial (ECI) frame and Barycentric frame, it is also consistent with the local cosmic microwave background (CMB) 'frames last scattered', and derives the effects postulated in the STR. Our ontological construction is again consistent, as it is with the finding from laser interferometry (Wise, Chhabildas 1985) that, in dynamic compression, refractive index data can '*exhibit a pronounced deviation*' from all predictions including the Lorentz-Lorentz and Drude relations. Kinetically, in the case of a mirror moving in a vacuum, the light arriving is transformed to c with respect to the mirror, and the light leaving the surface would be transformed back to the c of incident light, all within some ~10$^{-6}$m.

The Navier-Stokes hydrodynamic equations cannot be solved to within ~25% so precise maths gives way to small scale uncertainties. The kinetic relationship between each side of the 'collisionless' shock is then closely equivalent to the Fox (1965) view of velocity being localised to c by absorption and re-scattering via extinction and birefringence within the nanoscale shock. Doppler shift then initially arises from the ***real*** quantity of *wavelength* $\lambda$, modulating the time based derivation of frequency and conserving the constant c = f$\lambda$.

We must remember that to find any Doppler shift an observer must always *detect,* and use a different frame to the emitter. Being in a different frame **contravenes Proper Time rules**, so valid measurement of c and wavelength in the *emitter* frame cannot be made. Crossing the TZ boundary causes emitted wavelengths to contract or dilate, so *wavelength λ is not co-variant.* The TZ is then the spatial limit to the local inertial frame. Magnetohydrodynamics and re-emissions then create the boundary of a system's background 'state of motion', which may be termed the physical 'field limits' of the local inertial frame. $\Delta\lambda$ on transform then consistently derives *contraction and dilation of an event period.* What is more, shock ion state is uncountable, yet a 'group' motion is assignable. In this case *uncertainty* is only complexity and *remains entirely deterministic.* We suggest that perhaps the most compelling part of this dynamic hypothesis is that neither the component mechanisms or effects are unfalsifiable predictions but are already well known as atomic scattering and the co-variance of light speed. Both have strong experimental support, but the latter, as described in the second postulate of the STR, is as yet unexplained in quantum terms. No algorithmic basis has yet emerged to describe the physical position of a shock or TZ with respect to a massive body or system. We propose that the effects of influencing factors should be explored including, mass, relative motion in a local background, relevant wavelength and dominant em field. Ionospheric limit positions may initially only be established by observation, in an the same way as refractive index n. Spectroscopy and kinetic analysis of astronomical observation will become even more important tools at macroscopic scales. The transformation between inertial frames itself can, however, be described mathematically. The terms below are for an interacting observer, (detector), medium or plasma shock (n = 1) in relative motion v approaching the source.

| | | |
|---|---|---|
| Wavelength will be; | $\lambda_c = \lambda_{c+v} (1 + v/c)^{-1}$ | (1) |
| or for a receding observer; | $\lambda_c = \lambda_{c-v} (1 - v/c)^{-1}$ | (2) |
| The shifted wavelength is then; | $\lambda' = \lambda_O (1 + v/c)^{-1}$ | (3) |
| The shifted frequency; | $f' = f_O (1 + v/c).$ | (4) |
| Incident wave (not observable), at c is; | $\psi = \psi_O \sin 2\pi(f\,t + 1/_\lambda x)$ | (5) |

The wave equation remains invariant on transformation in Euclidean space, so the scattered wave is;

$$\psi' = \psi'_0 \sin 2\pi (f't' + {}^1/_{\lambda'} x')$$  (6)

Observed light speed, observer approaching source;

$$f'\lambda' = \{f(1 + {}^v/c)\}\{\lambda(1 + {}^v/c)^{-1}\} = f\lambda = c$$  (7)

Observed light speed, observer receding from source;

$$f'\lambda' = \{f(1 - {}^v/c)\}\{\lambda(1 - {}^v/c)^{-1}\} = f\lambda = c$$  (8)

The constant speed of light for observers in any state of motion interacting with all light signals is thus derived due to the fact that although c is co-variant, neither wavelength $\lambda$, or therefore its derivative frequency (f) are co-variant on transformation. The importance of considering observers 'class' and frame is highlighted: An observer *not* changing frame (so measuring indirectly) will *not* find *apparent* co-variant c as he is *not* then interacting with the original emission, but *will* find derivative f co-variant, so no Doppler shift.

## 10. Dynamic and Kinetic Logic

In the case of the B&B moving glass surface, the two kinetic systems of electrons mixed at the TZ would represent the inertial frames of the glass and of the quantum vacuum. If the velocity change is indeed at the TZ then a logical explanation exists for the change from the laws of geometrical optics to the unexplained non-linear effects and from Fresnel to Fraunhofer refraction at that zone. The same mechanism then resolves our central problem, namely the acceleration of the light leaving the glass plate to c with respect to the vacuum. The 'kinetic frame' of the vacuum emissions appears able to be defined by the incident light speed itself because photo-ionised electron flux in that frame increases with intensity and $\lambda$ decreases with increased incident velocity, thereby increasing frequency. The speed differential between the outer electron layer and the arriving light would need to be precisely c for any reciprocal emissions into the vacuum to also be at c. The *relative electron velocity across the TZ* would then be the co-variant frame speed v. The photoelectric effect on the outer electrons also emerges as incident photon and electron energy correlate. If one medium n value is above unity, normal non-kinetic refractive effects (2 and 3 in Part 3) may be implemented independently of the TZ.

We apply the modality principles of propositional dynamic logic (PDL) developed for computing but with broader relevance. At a quantum level (Baltag, Smets 2010) systems may be 'interleaved' but are kinetically separated. An acceleration is therefore required between leaves. We find this discrete 'planar' approach analogous to Einstein's concept of Cartesian co-ordinates as 'planes' "*rigidly attached*" to bodies. Motion is an invalid concept in geometry, and so limits or boundaries between geometries are implied, where one space time geometry is not valid in a different geometry. We suggest that Cartesian co-ordinates must then exist as 3D interleaved or 'nested' kinetic planes or bodies which do not co-exist spatially. If Einstein's smaller space s moves within S, it becomes bounded. The fundamental structure of truth function logic (TFL) is that a compound proposition may be part of another compound proposition, which may be part of another compound proposition ad infinitum. This precise nested hierarchical structure emerges, simply substituting kinetic states (frames) for propositions, showing the logical consistency of the model. The ontology implies that the similar spatial positions of Earth's shock and the *limits* of Earth's non rotating frame are **not** co-incidental. We propose a detailed analysis and possibly another experiment *[Exp.5]* involving orbiter or probe telemetry to investigate the relationship between the non-linear effects found in the region of the TZ and the effects of our planetary bow shock.

Invoking kinetic logic, and postulating that all detector lenses are equivalent to new media, we find that an approaching signal of any wavelength cannot hold any information as to the existence or motion of the ***new*** field or medium ***prior to interaction***. Logic then dictates that if signal velocity is altered **after** the initial interaction, then any quantum process involved would occur ***at*** that initial interaction point. Science may have assumed that a quantum mechanism is not needed, but Einstein was searching for just such mechanism to unify QM with the STR. The importance of direct ***interaction*** of observer with observed is confirmed, and consistent with ***the observer as part of*** the system. The proposed TZ mechanism may then also be considered as providing the Lorentz transformation (LT) by producing co-variance to the limit of gamma at OB electron density. The LT curve profile may prove to require a slight adjustment to align more closely with the 'input' power curve of LHC acceleration and the synchrotron emission frequency curve approaching c. The 'Unruh' (Crispino et al. 2008) effect would give similar photoelectrons, from speed *not* acceleration.

## 11. Michelson-Gale-Pearson (1925) Falsification.

Our hypothesis is kinetically similar to the Fresnel/Stokes/Heaviside 'dragged ether' theory, but with *particle interactions* modulating c. Planck dismissed the doubts Lorentz expressed about 'varying ether density' but the logical basis of Stokes thesis was unfalsified when the conception was forgotten with the rise of the STR. The suggestion of experimental falsification of 'ether drag' only really came with the Michelson-Gale-Pearson experiment (1924) with a 1.9km perimeter fixed mirror interferometer in a low vacuum. The result contradicted the original Michelson Morley 1887 (M&M) 'null' result, and was reported as fringe shifts due to ether flow with respect to the Earth's rotation in; "***stationary ether** as well as **in accordance with relativity**.*"! The mixing of ether and the SRT was highly problematic. The interpretation that 'ether drag' was 'disproved' is also logically inconsistent. The assumed basis of consistency with relativity is that rotation is a special case in the STR. The explanation of rotation within a 'fixed ether' refers to a fixed Earth centred local inertial (ECI) coordinate system, a concept similar to the surface ECRF but non rotating. The ECI frame then represents a 'dragged ether' frame. As the Earth's ionospheric shock does not rotate with the planet yet the atmosphere does, that ECI frame is consistent with our re-emission model at and beyond a planetary scale. The ECI is also the frame used by GPS, and both this frame and the ionosphere are 'at rest' only locally with Earth as we orbit the sun. Similarly the barycentric (Solar system) frame which each planet moves through is only at rest with respect to our own sun, extending only to the heliospheric bow shock and heliosheath at ~100 AU's. The group 'galactic arm' state of motion (frame) then applies. But where extinction distance is greater than shock thickness we will still find birefringence, scintillation and the non-zero results of Michelson, also those consistently reported by Dayton Miller always found greater at higher altitudes.

The arrangement of inertial frames in relative motion which emerges is precisely equivalent to Minkowski (1908) and Einstein's (1952) conception of;

"*endlessly many spaces in motion relatively.*" Their boundary is kinetic, or a 'change of assignable state of motion', and physically implemented by coupling and scattering in the diffuse medium in the same way that two moving bodies of water particles will always change light speed to the *local* c/n of the water. The kinetic interface of the moving bodies of water particles is analogous to local CMB scattering surfaces, explaining the CMB anisotropic velocities of various systems, and the frames last scattered of Scott & Smoot (2004). The process may be seen as '*continuous spontaneous localisation*' of c giving CSL.

Despite the above logical consistency, however, if the reported fringe shifts of the MGP experiment did occur due to Earth's rotation in an ether frame, then our local scattering basis may still be falsified at an atomic scale unless ether is invoked. This is because the mirrors are at rest in the incident frame, so light scattered at the surface of the mirror after reflection would then travel at c with respect to the mirror. A problem then exists. LIGO long baseline interferometry has not confirmed the 1924 MGP findings, and GPS one-way measurement (four orders more sensitive than the MGP experiment) also gives null ether flow results in the ECI system. We identify some possible systemic errors in the MGP experiment which may have caused the apparently anomalous result. Any asymmetry in a reciprocating beam system will produce a fringe shift. For instance any differences in the **number** of TZ shock crossings, surface encounters (of two TZ crossings each), reflections, passages through glass, or changes of media affect overall propagation time, so the relative propagation times, so changing the phase. The result is then in doubt and anomaly resolved.

The MGP set-up contained at least one of the above asymmetries, but the mirror construction is not described apart from being lightly (50% reflection) or heavily silvered (or gold). We cannot then identify the actual asymmetry present but the configuration cannot avoid at least one. The diagrammatic set up (MGP Fig.1) indicates four passages through mirror glass in each case, but also partly with 'incident surface' reflection and partly silvered back surface reflection (mirror A). If the diagram is correct in terms of reflective surface detail, the (delayed) counter clockwise beam has **eight** *changes* of medium (two each at *A,C,B* and *A*) and the clockwise beam only **six**, explaining it's earlier arrival. Reflection can polarise light and causes polarisation mode dispersion (PMD) propagation delays. If we then consider just shock crossings and reflections, the counter clockwise beam has one *additional* internal reflection (absorption and re-emission) requiring to be negotiated at mirror A.

If MGP's Fig. 1 simple 'surface' reflection was purely diagrammatic and more conventional back silvered mirrors were actually used, then subject to the set up of mirror *A* the counter clockwise beam may have encountered up to two additional glass transits. It is even conceivable that because some asymmetries are repeated in the shorter 'datum' path the fringes may have been misidentified on the initial '*correction'* of the '*apparent displacement,'* and the shift between each long and short path measured, although this might seem quite unlikely. The reason for the contradiction with the M&M null result is not discussed, but no other experiment has repeated the MGP finding. Our model is consistent with the M&M result and logical analysis by Christov (2006). If all systemic errors in the MGP experiment were eliminated, then ether would be required for our atomic scattering model, but otherwise the ontological foundation of our model is consistent with or without any 'sub-particle' medium.

## 12. Ontological Construction

We now propose and logically establish that there are *two* theoretical cases of observation and measurement of light speed which are allowed to give different results. We will term the first case '*direct*'; via the evidence of wavelength λ of directly received light, calculated from frequency *f* and assumed velocity using the constant c = *f* λ and energy conservation law e = *f* λ. We term the second case, '*indirect*'; via the evidence of light scattered from particles of a medium *charged by* the original pulse. Our proposed experiment [1] considers the medium as diffuse plasma in a cloud chamber. An observer *at rest* with the chamber may record a propagation speed c. But if he then passes by in a frame accelerated to v and records the apparent speed he may of course find ***apparent c + v***, (or c - v) *because the rules of Proper Time do not then apply*. If he is 'at rest' and the chamber passes by him, the result is the same. This is clearly because the light then observed is *only the light, **scattered** by the particles*. This light of course travels, locally, at c, but interpretation up to now has been that the original signal must violate the STR even if we measure *apparent* c + v!

Current mainstream interpretation commonly assumed that the 'light speed' derived from this scattering information cannot apparently exceed c from any observer frame. ***This assumption is incorrect*** and, we suggest is the cause of much confusion and questioning of the logic of the STR. We can only actually observe ***indirect*** light or em energy via atomic scattering however diffuse the medium particles. By definition then even the quantum vacuum is equivalent to a medium. *Apparent* c + v is then allowable in the case of *secondary* detection of *indirect* emissions but actual c remains constant ***within*** any frame and co-variant *between* frames, exposing the true intuitive logic of the STR postulates.

The concept of 'apparent' c+v turns out to be merely a re-interpretation of space-time consistent with the comment that; "...*cases with a velocity greater than that of light will henceforth play only some such part as that of figures with imaginary co-ordinates in geometry.*" (H. Minkowski 1908). We now consider a Minkowski space time 'event' (constituting any non-zero 'period' of time), negotiating a TZ. We find that the period of the event in the emitter frame is *changed* by the re-scattering mechanism at the TZ. The change in apparent 'period length' is entirely due to the wavelength change but co-variant propagation velocity c is derived. The apparent length of events will then not be co-variant but will be contracted or extended subject to v. Identical clocks will then run at the *same speed in all frames* but, quite intuitively, clocks in relative motion to any observer will simply appear to him to run at different speeds. A clock approaching an observer at 0.1c will then simply appear to run 10 per cent faster. Both SR and QM then use the same absolute time, but time period evidence, once emitted, does not commute between frames.

Photo-ionisation already tells us that incident light itself can produce the ions, and at rates subject to relative 'arrival' velocity so no 'ether' needs to be invoked for the process of light re-emission. This removes the apparent inconsistency of the concept of CMB 'frames last scattered' (Scott, Smoot 2004) with the SR postulates. The varying CMB anisotropic flows found are real, with real ion shock 'scattering surface' boundaries between them. The quantum vacuum and interstellar medium are endowed by this 'discrete fields' model (DFM) with the qualities of a very diffuse medium, condensing to a local modulating plasma medium with any motion of mass within it. Local dynamic background inertial frames result, with speed always limited only by the local background defined by arriving light speed. A scale invariant process implies that infinitely many such 'nested' states of motion would exist. The speed of Earth would then bear no direct relationship to the speed of a distant star, because Proper Time ***does not apply***. Logical equivalence is found with the Minkowski (1908) and Einstein (1952) conception of; "*endlessly many spaces in motion relatively."* The fine structure constant of $1/137^{th}$ may perhaps also be considered as a TZ 'ground state', at rest. The proposed experiment under varying laser light intensities and in a dynamic mode would falsify this suggestion with regard to kinetic variation. An additional laser experiment is proposed where a secondary signal beam is harmonically merged with and carried by an intense principal beam *[Exp.6]*. Signal arrival times are predicted to vary at certain harmonics, somewhat similar to 'quantum tunnelling' effects, and apparent superluminal pulses commonly found in quasar jets such as those of M87 *(Add.; Rees, 1985).*

## 13. Theoretical constraints

We have found that all 'Proper' speed is **propagation speed.** Solutions based on deterministic Local Reality have been constrained by Bell's, Haag's and Currie-Jordan-Sudarshan (CJS) theorems. Haag disallowed any field theory requiring 'invariance'. As λ is indeed not 'co-variant' on transformation between frames the terms are met. CJS states that two frames can't interact or exchange energy for the LT to apply. But our underlying mechanism when approaching plasma optical breakdown limit gamma now produces the LT in the case of observation by *direct* interaction. In the *indirect* case 'proper time' *does not apply* so apparent c+v is allowed. Bell's theorem is agreed but uses incomplete starting assumptions. Von Neumann showed that consistent QM must have uncertainty at each detector. A higher order hierarchical dimension or 'sample space' equivalent to a Bayesian or Godel n-value ('fuzzy logic') quantum probability distribution ***between*** the cardinal values 0 and 1 then offers resolution of the EPR paradox, by decoding the 'noise' limiting the Shannon channel capacity. This may be seen as adding a $3^{rd}$ dimension to a cosine wave form to create a helix over time, using orbital angular momentum as the 'real' energy path of an entangled particle. We predict then that in time resolved single pair analysis of entanglement experiments an orbital asymmetry will be found which is not accessible via statistical (weak measurement) 'correlation' techniques.

This model would give a locally 'semi'-deterministic case which has analogies with virial 'inertial systems' and 'quantum mixing' ('t Hooft 2009). The field based ontology invoked then avoids the previous constraints on Local Reality in the unified field solution as sought by Einstein, but with a reducing layered uncertainty of an evolving fractal nature. Parallels with these hierarchical 'gauges' or groups and Kaluza 'dimensions' then exist, which we suggest may then have an ultimate limit at gamma for condensed matter.

The function of the hyperbole of 'rapidity' in constraining apparent c+v is no longer required, so it's additive qualities can be returned as the '*imaginary*' c+v of Minkowski. The new understanding that apparent c+v is allowed because it is an optical illusion from indirect emissions may be the most (apparently) surprising part of the DFM ontology but becomes an inevitable logical truth once the detection mechanism is understood. 'Rapidity' may then become a more useful word, to distinguish real local ('proper') propagation velocity (<c) from the apparent 'velocities' derived only via sequential scattering evidence.
*(Addendum; Online version only; Mechanism and implications of the above including a helical 'IQbit' are expanded on in; Jackson 2013).*

## 14. Conclusions
We have found and described a consistent resolution to the B&B anomalies via an ontological construction of known quantum mechanisms and in accordance with the postulates of Einstein's Special Theory of Relativity. We have also, however, shown that the conflicting result of Kantor's experiment may have been due to the air flow along the beam paths caused by the rotating plates. We find similarity with Fox's (1965) localisation of c via extinction, and *apparently* successful refutations of that thesis for vacua, including by Brecher (1977). We have further theoretically investigated mechanisms at the emitter surface, more consistently explaining Maxwell's near field transition zone, plasmons and non-linear optics. We refer to apparently anomalous and related astrophysical findings, and also identify and discuss a connection with the kinetic reverse refraction effect of geometrical optics, equivalent to the breakdown of linear optics and Fresnel refraction at the Fraunhofer distance. The rotation of optical axis giving refraction alongside Faraday rotation of polarity is found to derive elliptical polarity and stellar aberration while conserving causal wavefronts by showing the optical axis is ***not*** tied to the wavefront normal (as Calcite crystals reveal). This significant stellar aberration component is as added by hand by the US Naval Observatory to get the accurate predictions of the AA2010 data. The missing underlying relativistic theoretical basis for kinetic refraction in frame transition is identified in the ontology as arising from charge asymmetry.

We propose a possible new set of consistent kinetic relationships and quantum mechanisms arising from photo-ionization, Raman scattering, extinction and many other familiar phenomena at the refractive plane of co-moving media. Previously assumed interpretations of classical theories are falsified with the application of the modality of propositional dynamic logic as developed for computing. We find the conceptual 'interleaved' basis of modal logic applies perfectly to Einstein's and Minkowski's conceptions. Cartesian co-ordinate systems as within space-time geometries are not required to move, but each geometric system itself may move and has a discrete domain limited by acceleration. Speed is only then relevant to the local background, yet is set by atomic scattering from matter, constantly modulating light propagation at the near field limit. Two distinct 'speed change' components are then identified; the first caused by refractive index n from interaction time, the second kinetic caused by the relative motion of the re-emitting particle compared to the previous emitter. Both contribute to wavelength change so also derived Doppler shift of observed frequency. The effects of lateral motion are aberration and birefringence found as rotation of re-emitted optical axis due to the asymmetry of charge the motion brings, giving the kinetic reverse refraction.

We have identified that constraints such as Bells inequality, CJS and Haag's theorem are overcome in our proposed ontology by the 'detection' change to λ. We have discussed the 1924 Michelson-Gale-Pearson (MGP) interferometer experiment, which conflicts with the 1887 M&M null result, and has not been confirmed. The MGP result is the only cited falsification of the 'ether drag' theory, kinetically similar to our discrete field model. We have logically analysed the MGP experiment and shown how the result would be consistent with the non-rotating ECI frame, but not with an 'ether flow' caused by our orbital path. In this case the result, though proving little, does not disprove our ontology of kinetic 'nesting'. We do however also find and explain a systemic error in the MGP experiment which Michelson would not have known of in 1924. There is an asymmetry of paths, with the anticlockwise path undergoing either an additional reflection or two additional glass transits (subject to mirror configuration). Either would give a fringe shift and confuse main and 'control' beam fringe identification. We have also shown the logical equivalence with GPS, and with CMB anisotropy, where the concept of CMB 'frames last scattered' is confirmed as consistent with the STR postulates. The domain boundaries and the mechanism are identified, with scattering across the full complex plasma frequency range giving continuous spontaneous localisation (= CSL). Modelling may best use 3D Lagrangian Coherent Structures.

Intriguing possibilities arise from the relationship of the quantum mechanisms discussed and resultant effects consistent with the postulates of the STR. All inconsistencies of SR and QM evaporate. We identify and propose possible experiments to falsify the theory (see below). We find that using the quantum mechanisms invoked, the wave equation is co-variant on transformation, but wave-length lambda is not. Two **classes** of observer frame are identified and the different results are found as critical. Speed may be *real* (direct interaction) or *apparent* (indirect observation). For secondary scattering detections from motion in *another frame* the Proper Time rules do not apply, and the *apparent* delta c and *apparent* c + v are allowed, balanced by *apparent* d/dt and delta f to maintain the constant; c = f'λ. The Laws of conservation are thus satisfied.

**Suggested Experiments;**

*[Exp.1]* Plasma chamber interferometer and re-emitter delta v, to compare optical axis path arrival times. p2.

*[Exp.2]* A partially evacuated chamber at varying densities to compare n, extinction and birefringence. p2.

*[Exp.3]* Varying surface transitions for a given glass thickness. We predict the B&B. +0.01 fringe shift. p3.

*[Exp.4]* B&B vacuum chamber with varying intensity lasers to quantify the kinetic effects on TZ activity. p6.

*[Exp.5]* Orbiter or probe telemetry to allow analysis of our planetary bow shock for non-linear TZ effects. p8.

*[Exp.6]* Time resolved single photon entanglement experiment and analysis to find a quantum cosine distribution independently produced at each detector.